**An inverse analysis of cohesive zone model parameter values for human fibrous cap mode I tearing**

Xiaochang Leng[1], Lindsey Davis[3], Xiaomin Deng[1], Tarek Shazly[2], Michael A. Sutton[1], Susan M. Lessner[3]

[1]Institute of Engineering Mechanics, Nanchang University,
 Jiangxi, 330031, People's Republic of China

[2]College of Engineering and Computing, Biomedical Engineering Program,
University of South Carolina
Columbia, SC 29208

[3]School of medicine, Department of Cell Biology & Anatomy
University of South Carolina
Columbia, SC 29208

**Corresponding author:**
Xiaochang Leng, Ph.D.
Email: lengxc1984@163.com
Address: 999 University Avenue, Honggutan District, Nanchang,
Jiangxi, 330031, People's Republic of China
Tel: +86-0791-83969641

**ABSTRACT**

Atherosclerotic plaque failure results from various pathophysiological events, with the existence of fibrous cap mode I tearing in the arterial wall, having the potential to block the aortic lumen and correspondingly induce serious clinical conditions. The aim of this study was to quantify the interfacial strength and critical energy release rate of the fibrous tissue across the thickness. In this study，an inverse analysis method via finite element modeling and simulation approach was presented. A cohesive zone model (CZM) was applied to simulate the tearing of the fibrous cap tissue under uniaxial tensile tests along the circumferential direction. A fiber-reinforced hyperelastic model (Holzapfel-Gasser-Ogden) was implemented for characterizing the mechanical properties of bulk material. With the material parameter values of HGO model from inverse analysis process as the input for the bulk material, the interfacial strength and critical energy release rate along the tearing path or failure zones are obtained through the same method as material identification process of HGO model. Results of this study demonstrate the fibrous cap tissue tearing failure processes.

# 1. INTRODUCTION

Atherosclerotic plaque rupture is a serious and complicated event, which often leading to severe cardiovascular diseases such as myocardial infarction or stroke [1, 2]. The plaque rupture may occur at the interface between plaque and underlying medial layer [3] or the interface between fibrous cap and underlying plaque tissue [4]. Fibrous cap tearing under mode I condition is one way that prone to induce the plaque rupture. In order to restore the lumen diameter of carotid artery with atherosclerotic plaque, implementation of balloon angioplasty or stenting may induce extension of the shoulder or central area of fibrous cap, triggering tearing of fibrous cap across the thickness [5-7]. The plaque failure phenomenon is a very slow process. At the beginning, degradation of fibrous cap tissue accumulates with the cyclic inherent stress from the blood or stenting implementation on the lumen of artery, which cause micro tearing propagation and stress concentration inside the fibrous cap [8]. This phenomenon initiates and accumulates mechanical damage, gradually leading to the macro tearing across the thickness of fibrous cap. At last, the atherosclerotic plaque is prone to rupture when the damage reaches a threshold level [9]. Thus, the material properties, interfacial strength and critical energy release rate across the thickness of the fibrous cap play an import part in the atherosclerotic plaque rupture [3, 4].

Numerous studies have evaluated the mechanical or failure behavior of fibrous cap or plaque. The ultimate tensile stress and strain of human carotid artery atherosclerotic plaques which contain lipid cores, intra-plaque hemorrhage and a thin fibrous cap were obtained through the uniaxial tensile test [10]. The shear modulus of the human fibrous cap was quantified via indentation tests [11]. The circumferential tensile stress and associated strain of human fibrous caps were obtained through the uniaxial tensile tests [12-15]. The ultimate tensile stress and strain are the material parameters used to evaluate the breakage of the fibrous cap tissue [16], but those parameters cannot be used to assess the damage initiation and propagation processes.

Thus, the cohesive zone parameters such as interfacial strength and critical energy release rate can be implemented as an input data in the finite element model to simulate and predict the fibrous cap failure process.

The critical energy release rate can be obtained from the plaque delamination [17] and fibrous cap delamination tests with the delaminated areas acquired directly from the experiments. The method cannot be used directly to acquire the adhesive strength of the fibrous cap across the thickness with the limit information of the teared area from the tearing experiments. Thus, an automatic inverse modeling approach via finite element analysis method provides a way to quantify the interfacial strength and critical energy release rate from the fibrous cap tearing tests.

Numerous automatic methods have been used to obtain the material parameter values of arterial tissue under different types of mechanical tests. An inverse boundary value method with finite element simulations and curve fitting tool from Matlab was implemented to acquire the shear modulus of human fibrous cap under indentation tests [11] and material parameters of anisotropic constitutive model on mitral valve anterior leaflet [18]. In other studies, an inverse optimization method based on Nelder-Mead Simplex method and FEA method was conducted to estimate in vivo material parameters for a human aorta [19] and the material parameters of human carotid arteries were determined by Levenberg-Maquardt algorithm [20-24].

So far, little information about the interfacial strength and critical energy release rate of interface across the fibrous cap has been found. In this study, a mode I tearing test on human fibrous cap specimen from circumferential direction was carried out in order to obtain the adhesive strength across the thickness using inverse analysis method. The finite element analyses of the tearing procedure were proposed, providing a comprehensive understanding of the fibrous cap tearing phenomenon.

## 2. MATERIALS AND METHODS

### 2.1 Experimental procedure

In this study, fibrous cap tearing experiments were performed on human carotid artery obtained from endarterectomies. Five fresh carotid endarterectomy samples were obtained at the time of surgery from five patients and one specimen from each sample was successfully prepared, providing samples for five fibrous cap tearing (FCT) tests. Those fibrous caps were cut along circumferential direction to yield strip specimens.

Prior to the tearing test, samples were preconditioned by 5 cycles of quasi-static uniaxial tensile tests with loading rate 0.05 mm/s in order to get repeatable mechanical response. These obtained experimental data was used to identify the bulk material properties of fibrous cap. Then, a surgical scissors was used to carefully introduce an initial cut perpendicular to the loading direction (keep the tearing path in a straight line) at the midline of the specimen (Fig. 3a, c). The lower edge was gripped by the clamps connected to the load cell of a test system (Bose ELF 3200, Biodynamic Co, MN) for load data recording, and the top edge (proximal end) of the fibrous cap was gripped by the clamps connected to actuator which controls the applied loading (Bose ELF 3200, Biodynamic Co, MN ) [25].

The tearing test was performed by moving the proximal end of the fibrous cap at a loading rate of 0.05 mm/s. The prescribed displacement and resultant load were recorded via system actuator and load cell. The tearing process was captured by a computer vision system with two cameras perpendicularly positioned so as to get the front and side views of the specimen.

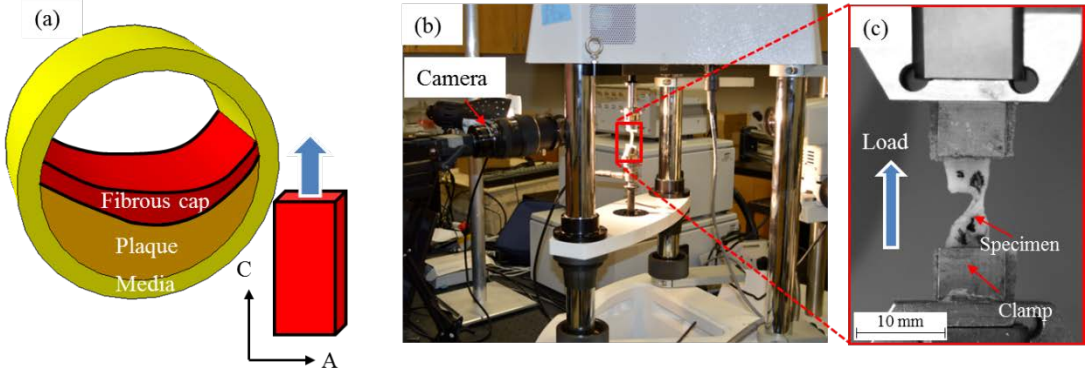

**Fig 1.** Schematic of experimental setup. (a) Schematic of carotid artery plaque specimens obtained from endarterectomies and cut to strips transversely (A: axial direction; C: circumferential direction); (b) Experimental setup of fibrous cap tearing tests; (c) experiment setup (zoomed-in view).

## 3. Theoretical Framework

### 3.1 Bulk material model and interface damage for porcine aorta

For the simulation of human fibrous cap tearing, the Holzapfel-Gasser-Ogden (HGO) model [26-28] and cohesive zone model (CZM) [29, 30] were adopted for the bulk material behavior and the tearing behavior along the interface in the fibrous cap layer, respectively.

### 3.1.1 Bulk material model

The HGO model characterizes the fiber-reinforced behavior of arterial tissue. The collagen fibers are assumed to distribute parallel to the circumferential direction of aorta with certain angle. Thus, the strain energy potential for the arterial tissue is expressed in a decoupled form as

$$\Psi = \Psi_{vol} + \bar{\Psi} \tag{1}$$

The volumetric part, $\Psi_{vol}$, is given by [31, 32]

$$\Psi_{vol} = \frac{1}{D}\left(\frac{J^2-1}{2} - \ln J\right) \tag{2}$$

where $\frac{1}{D}$ represents the bulk modulus of the material.

Hence, the free-energy function $\overline{\Psi}$ [33-35], is expressed as

$$\overline{\Psi}(\overline{C}, H_1, H_2) =$$

$$\frac{\mu}{2}(\overline{I}_1 - 3) + \frac{k_1}{2k_2}\left[e^{k_2[\kappa\overline{I}_1 + (1-3\kappa)\overline{I}_{41}-1]^2} - 1\right] + \frac{k_1}{2k_2}\left[e^{k_2[\kappa\overline{I}_1 + (1-3\kappa)\overline{I}_{42}-1]^2} - 1\right] \qquad (3)$$

where $\overline{I}_1 = tr(\overline{C})$ is the first invariant of $\overline{C}$, and $\mu$ is the shear modulus of the material without fibers; $\overline{I}_{41}$ and $\overline{I}_{42}$ are tensor invariants equal to the square of the stretch in the direction of two families of fibers, respectively**Error! Bookmark not defined.**; the constitutive parameter $k_1$ denotes the relative stiffness of fibers; $k_2$ is a dimensionless stiffness [36-38]; $\kappa$ is the dispersion parameter which have to important values that one family of fibers are parallel to each other when $\kappa = 0$ or distribute isotropically when $\kappa = 1/3$.

### 3.2 Interface damage model

An irreversible exponential CZM with normalized effective traction and effective displacement is shown in Fig. 2, the local effective traction considering the loading conditions have the expression as [39-41]

$$t = e\sigma_c \frac{\delta}{\delta_c} exp(-\frac{\delta}{\delta_c}) \quad \text{if } \delta \geq \delta_{max} \text{ or } \dot{\delta} \geq 0 \qquad (4)$$

where $\sigma_c$ is the strength of the cohesive interface and $e = \exp(1) \approx 2.71828$; $\delta_c = \frac{G_c}{e\sigma_c}$ denotes the effective displacement when $t = \sigma_c$; $\delta_{sep}$ is the effective displacement when the cohesive interface element damage completely and $K$ is a penalty stiffness.

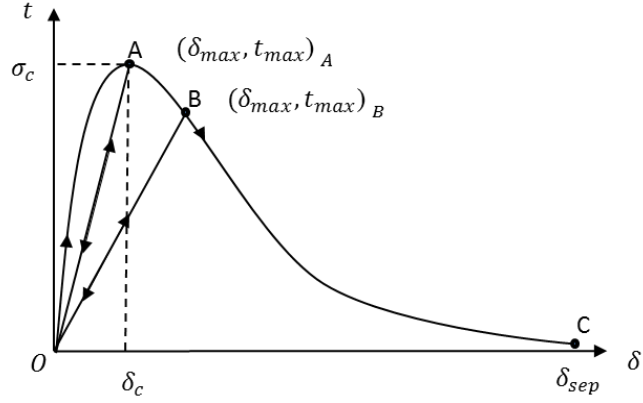

**Fig 2.** Irreversible exponential cohesive model denotes with normalized effective traction and effective displacement.

When the unloading process occurs, the traction can be expressed as

$$t = \left(\frac{t_{max}}{\delta_{max}}\right)\delta \quad \text{if} \quad \delta < \delta_{max} \text{ or } \dot{\delta} < 0 \tag{5}$$

The work of separation per unit cohesive surface has the forms:

$$G_c = e\sigma_c\delta_c \tag{6}$$

The irreversible mechanism represents the damage behavior of the cohesive element. Thus, a damage parameter is introduced to characterize the damage of the cohesive interface during the loading and unloading conditions, which has the form as

$$d = \frac{e\sigma_c\delta_c\left[1-\left(1+\frac{\delta_{max}}{\delta_c}\right)exp(-\frac{\delta_{max}}{\delta_c})\right]}{G_c} \tag{7}$$

The damage parameter $d$ ranges from 0 to 1, which represents no damage exists in cohesive surface when $d = 0$ or a completely damage of the cohesive surface when $d = 1$. The damage accumulates during the loading process and the traction-separation relation will go along line AO or BO when unloading happens at point A or B. The CZM model was implemented in the commercial finite element code ABAQUS through a user UEL subroutine.

### 3.3 Inverse analysis method

The inverse analysis method is an automatic process to obtain a set of parameter values through fitting to the experimental data [42-44]. The objective function describes the difference between the predicted and experimental results, is defined to be

$$f = \sum_{i=1}^{n}\left[F_{p_i} - F_{e_i}\right]^2 \tag{8}$$

where $F_{p_i}$ are the predicted results (forces), which have two expressions as $F_{p_i}(\mu, k_1, k_2, \gamma, \kappa,)$ when the inverse analysis of material properties of HGO model is conducted and $F_{p_i}(G_{c1}, \sigma_{c1}, \dots G_{cm}, \sigma_{cm})$ ($m$ denotes the groups of adhesive strength distributing along the tearing path) when the inverse analysis of interfacial strength and critical energy release rate are calculated. $F_{e_i}$ are experimental results (forces), at the $i$th increment. A reasonable set of parameters values would yield when the objective function is minimized to an acceptable value that the optimization procedure stopped when the variation of the objective function is less than the set tolerance (TOL $= 10^{-6}$).

### 4. Numerical Implementation

The simulations of fibrous cap tearing process were implemented through the finite element analysis (ABAQUS 6.13, SIMULIA Inc) [31]. The material parameter values of the bulk material of fibrous cap were identified from the inverse analysis of the uniaxial tensile tests of fibrous cap before the tearing tests. Those material properties were used as the input data for the bulk material of the simulation of fibrous cap tearing procedure. Finally, the adhesive strength including the interfacial strength and critical energy release rate were obtained through the inverse analysis of the fibrous cap tearing tests using CZM-based approach by minimizing the difference between the numerical predictions of load-displacement curves and the experimental measurements.

## 4.1 Geometrical modeling

The geometrical models were reconstructed from the images captured during the fibrous cap tearing experiments (Fig. 3). The tearing path of the tests was determined directly from the experimental images taken at different stages during the testing. For specimen #1 (FCT1), the fibrous cap was totally tore into two separated parts, thus the tearing process was used as a representative illustration of the inverse analysis method to obtain the adhesive strength of the fibrous cap layer. For the rest of specimens, only part of the layers was tore, leaving approximately 1/4 length of the width of specimen stretched without separation. Thus, a straight tearing path was pre-defined along the width of the specimen at the beginning of the simulations.

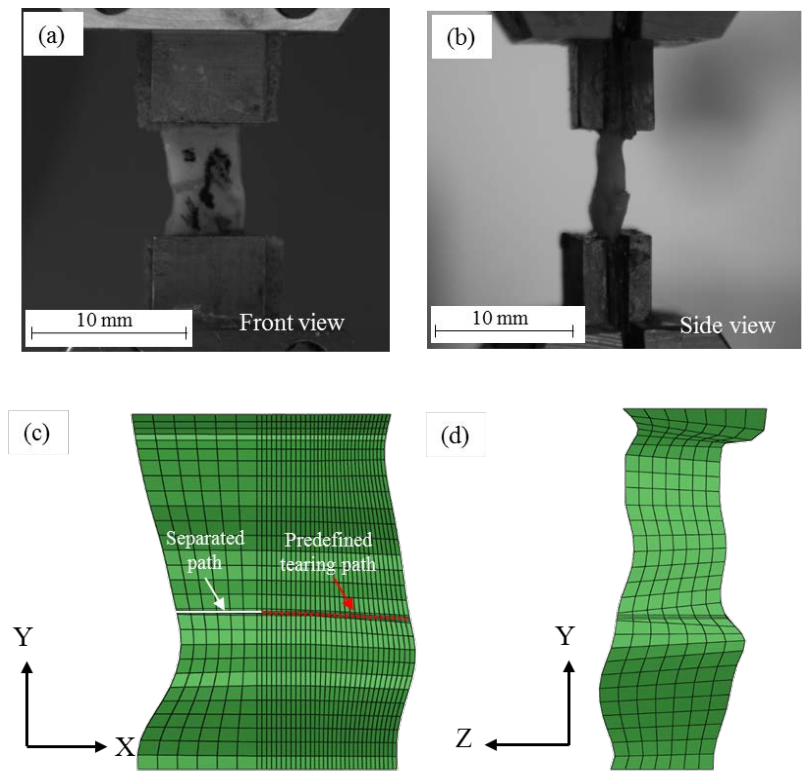

**Fig 3.** Representative experimental images of the fibrous cap tearing specimen: (a) a front view of the specimen; (b) a side view of the specimen; (c) FE model of front view of the specimen, which the white line shows the initial separated path and the red dot line denotes the predefined tearing path that will tear during the experiment; (d) FE model of side view of the specimen.

## 4.2 Meshing

The fibrous cap was meshed with eight-node brick elements (C3D8H), while the cohesive interface (tearing path) was meshed with zero thickness eight-node 3D user-defined elements which placed along the tearing path starting from the initial tearing front to the end of the sample along the width. The meshed geometrical model of FCT1 is shown in Fig. 3c, d, with 0.4 mm for the global size of the elements of fibrous cap and 0.1 mm for the cohesive element.

## 4.3 Boundary conditions

At the beginning of the experiment, the lower edge was fixed by clamps. Therefore, in the finite element model the boundary condition for the lower edge was taken to be that all nodes on the low edge were fixed so that they could not move in any direction. The top edge was fixed by clamps which were translated with a certain displacement upward with loading rate 0.05 mm/s, as shown in Fig.1c. Except for the fixed part of the two edges, all other surfaces of the finite element model were set to a traction-free boundary condition.

## 4.4 Identification of HGO model parameter values

The material parameter values of HGO model were obtained through inverse analysis of the uniaxial tensile tests of fibrous cap that fitting simulation predictions of the load vs. load-point displacement curves with experimental measurements (Fig.

4). A set of HGO parameter values of fibrous cap from the reference was set as the initial input values, a fully automated inverse procedure [45, 46] was adopted and all parameter values were determined when the objective function $F_{p_i}(\mu, k_1, k_2, \gamma, \kappa,)$ was minimized to an acceptable value from the numerical identification procedure.

### 4.5 CZM material properties along the tearing path

The HGO model parameter values obtained from the material parameter identification procedure were used as the material properties of bulk material. A set of exponential CZM parameter values obtained from the reference [16] was used as the initial guess of CZM parameter values for the cohesive interface. For the sake of simplification, the value of K and $\lambda$ were set equal to 10 N/mm$^3$ and 1 [16], separately. The initial guess of the critical energy release rate and interfacial strength for the total tearing path were 0.23 N/mm and 0.2 MPa (assumed value), respectively. For all the tearing path, only the values of critical energy release rate and cohesive interfacial strength varied during the inverse analysis process and the parameter values were determined when the objective function $F_{p_i}(G_{c1}, \sigma_{c1}, \dots G_{cm}, \sigma_{cm})$ was minimized to an acceptable value. The whole inverse modeling of fibrous cap tearing tests are shown in Fig. 4.

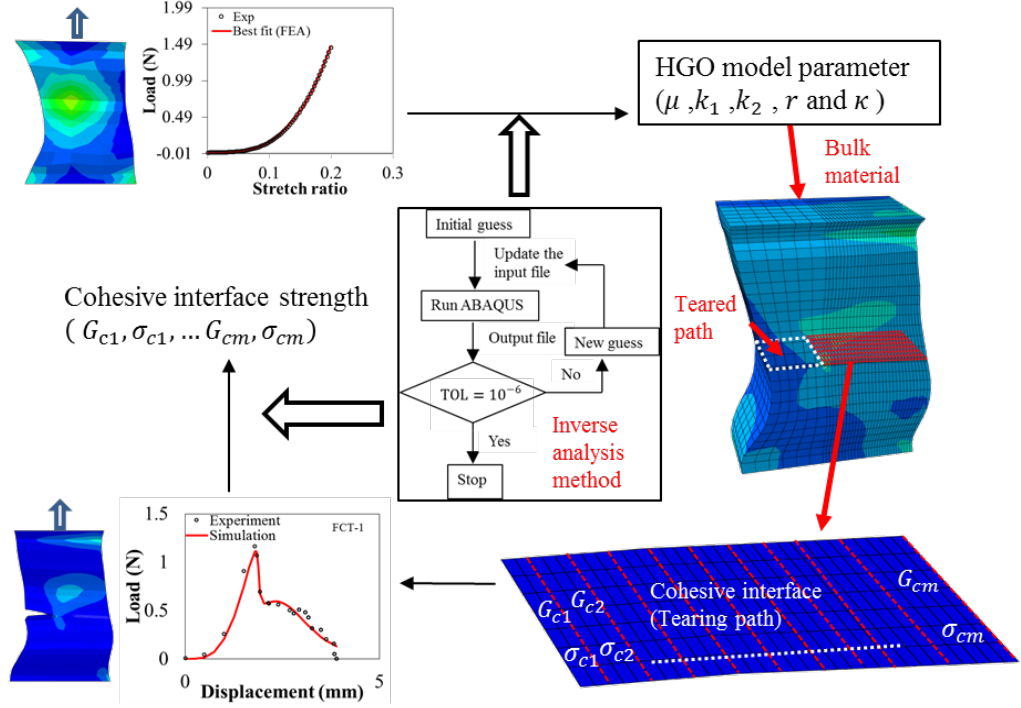

**Fig 4.** Flow chart of the inverse analysis of fibrous cap tearing tests which used to obtain the cohesive interface parameters along the tearing path.

# 5. RESULTS

## 5.1 Inverse analyses of uniaxial tensile tests of fibrous cap

In the numerical identification process, the HGO model parameter values of human fibrous cap were obtained when the objective function achieved an acceptable value, which are shown in Table 1.

**Table 1**. Best-fit parameters for the HGO model of human fibrous cap samples under uniaxial tensile tests.

| Sample | $\mu$ (kPa) | $k_1$ (kPa) | $k_2$ | $\gamma$ (angle) | $\kappa$ | Residual |
|--------|-------------|-------------|-------|------------------|----------|----------|
| FCT1 | 2.00 | 4705.30 | 54.26 | 57.01 | 0.27 | 0.0006642 |
| FCT2 | 2.00 | 413.9 | 37.96 | 66.69 | 0.27 | 0.0000013 |
| FCT3 | 5.1 | 938.80 | 20.09 | 38.53 | 0.23 | 0.0000225 |
| FCT4 | 4.00 | 388.9 | 16.26 | 65.87 | 0.25 | 0.0000258 |
| FCT5 | 4.40 | 3263 | 77.83 | 73.56 | 0.29 | 0.0000056 |

## 5.2 Inverse analyses of fibrous cap tearing experiments

In the previous section, a finite element model was developed for the strip of fibrous cap uniaxial tensile experiment and the parameter values for the bulk models were identified. In this section, the HGO model parameter values was used for the bulk material model of the fibrous cap in order to obtained the cohesive interface parameter values along the tearing path (Fig. 4). The values of critical energy release rate and CZ interface strength of fibrous cap are shown in Table 2. The comparison of simulation predictions between experimental measurements of the load-displacement curves of the fibrous cap tearing tests are shown in Fig. 5.

**Table 2.** The values of critical energy release rate and interfacial strength of fibrous cap tissue

| | FCT-1 | | FCT-2 | | FCT-3 | | FCT-4 | | FCT-5 | |
|---|---|---|---|---|---|---|---|---|---|---|
| Area | $G_c$ (N/mm) | $\sigma_c$ (Mpa) | $G_c$ (N/mm) | $\sigma_c$ (Mpa) | $G_c$ (N/mm) | $\sigma_c$ (Mpa) | $G_c$ (N/mm) | $\sigma_c$ (Mpa) | $G_c$ (N/mm) | $\sigma_c$ (Mpa) |
| 1 | 0.237 | 0.368 | 1.341 | 0.572 | 0.089 | 0.029 | 0.797 | 0.289 | 0.172 | 0.026 |
| 2 | 0.225 | 0.387 | 1.346 | 0.483 | 0.090 | 0.042 | 0.786 | 0.339 | 0.095 | 0.001 |
| 3 | 0.203 | 0.378 | 1.391 | 0.492 | 0.110 | 0.051 | 0.779 | 0.518 | 0.213 | 0.096 |
| 4 | 0.195 | 0.378 | 1.428 | 0.506 | 0.123 | 0.067 | 0.777 | 0.558 | 0.289 | 0.197 |
| 5 | 0.434 | 0.211 | 1.313 | 0.494 | 0.085 | 0.093 | 0.778 | 0.564 | 0.182 | 0.354 |
| 6 | 0.411 | 0.195 | 1.235 | 0.474 | 0.125 | 0.120 | 0.778 | 0.543 | 0.257 | 0.392 |
| 7 | 0.411 | 0.195 | 1.224 | 0.500 | - | - | - | - | 0.224 | 0.386 |
| 8 | 0.413 | 0.198 | - | - | - | - | - | - | - | - |
| 9 | 0.341 | 0.159 | - | - | - | - | - | - | - | - |
| 10 | 0.188 | 0.082 | - | - | - | - | - | - | - | - |
| Mean | 0.306 | 0.255 | 1.326 | 0.503 | 0.096 | 0.057 | 0.782 | 0.468 | 0.205 | 0.207 |
| SD | 0.105 | 0.112 | 0.075 | 0.032 | 0.026 | 0.040 | 0.008 | 0.122 | 0.063 | 0.171 |
| $f$ | 0.119 | | 0.021 | | 0.005 | | 0.078 | | 0.007 | |

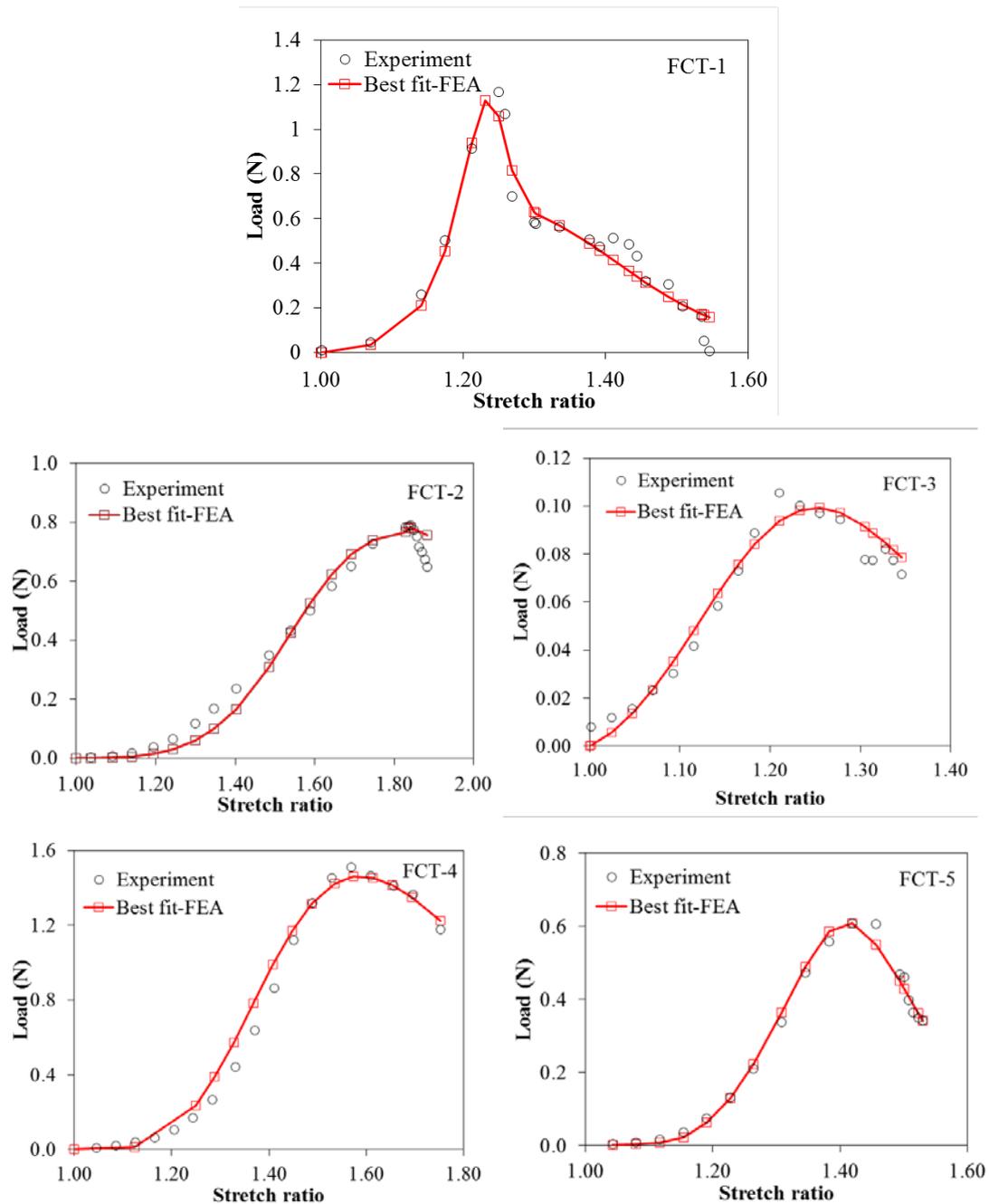

**Fig 5.** The simulation predicted load-displacement curves of loading-tearing cycles are compared with the experimental measured curves.

To obtain a quantitative understanding of the inverse iteration procedure, Fig. 6a shows the variations of CZM parameters (only the parameter values of the areas from the initial tearing front, the middle tearing and the end of the tearing path) with iteration numbers of inverse modeling of tearing simulation of sample FCT-1. It is noted that the predicted interfacial strength and critical energy release rate converged to certain values after approximately 30 iterations. The convergence trend of the objective function was shown in Fig. b, and the value was quickly reduced from a large value to a small value after seven iterations.

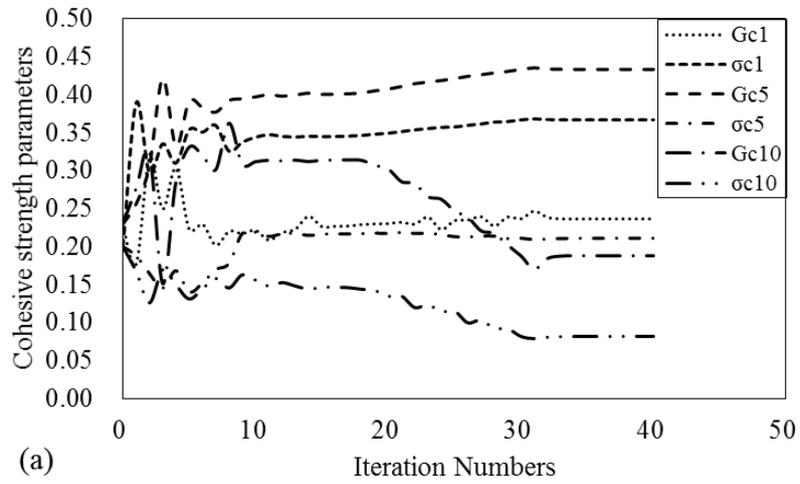

(a)

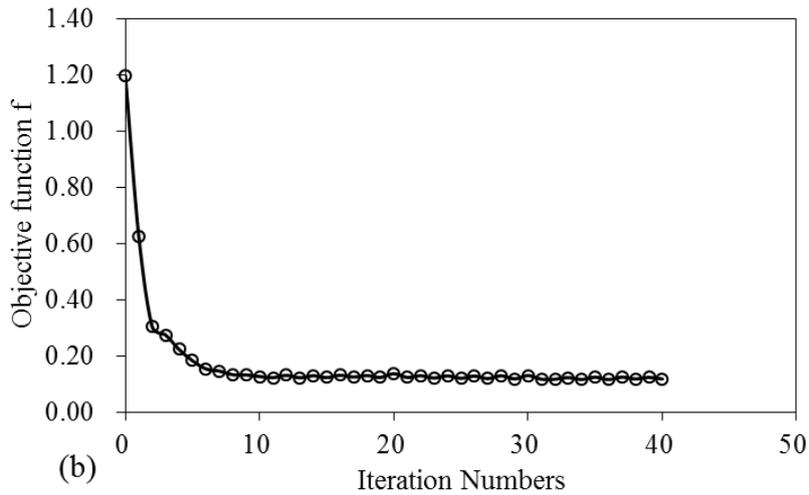

(b)

**Fig 6.** (a) Variations of CZM parameters and (b) objective function with iteration numbers of inverse modeling of tearing simulation of sample FCT-1.

## 5.3 Finite element results of the fibrous cap tearing process

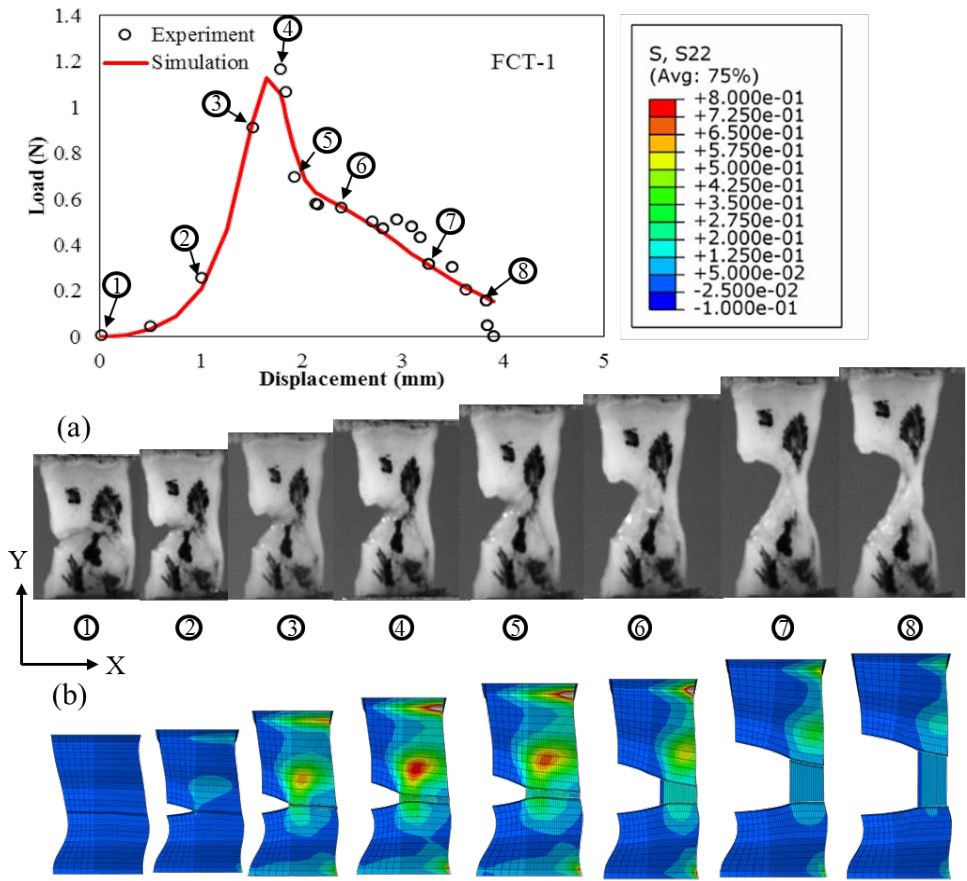

**Fig 7.** Tension stress (S22, along Y direction) contours at eight typical points along the load-displacement curve during tearing process: (a) tension deformation from experiment and (b) simulation prediction.

The predicted tension stress (S22) contours for eight typical points along the loading-displacement curve during the tearing process from the simulation are shown in Fig. 7. The deformation states from experiment (Fig. 7a) and the tension stress contour levels (Fig. 7b) are consistent with the corresponding loading levels in the load-displacement diagram. At beginning of the tearing test (Fig. 7a, at point 1), the sample deformed (Fig. 7a, b, at points 2 and 3) with extension along Y direction to the time point when the maximum tension stress occurs (Fig. 7a, b, at point 4). The loading level at point 4 was the highest and is sufficient to grow the tearing, and the resulting tension stress contour showed that the highest tension stress occurs in the cohesive interface and also the bulk material part. After that, the tension stress decreased with the tearing propagating along the tearing path (Fig. 7a, b, at points 5 and 6). At the loading point 7, the tearing propagation stops and the rest part of the tearing path (cohesive interface elements) deform with large extension which was similar to the uniaxial tension test. At last, the tearing begins propagating when resistant force decreases (Fig. 7a, b, at point 8).

## 6. DISCUSSION

To identify HGO model parameter values, uniaxial tensile tests on five fibrous cap samples were used. Since the mechanical properties of arterial tissue play an important role in identifying the critical energy release rate and the interfacial strength across the thickness of fibrous cap, the HGO parameter values were obtained through the inverse modeling method. The shear modulus of the matrix material μ were in the range from 1 kPa to 2.55 kPa, whereas the values in the other reference were 43.78 kPa [47] and 24.12 kPa [48]. However, the fiber reinforced mechanical response of arterial tissue is associated with the constitutive parameters $k_1$ and $k_2$, which dominate the mechanical behavior and the parameter values are in the same range as those in the reference [47, 48]

For the simulation of arterial tissue delamination, the ultimate tensile stress has been used as the cohesive interfacial strength for the CZM [49-51]. In this study, it was found that the cohesive interfacial strength of the fibrous cap along circumferential direction was in the range from 0.029 to 0.572 MPa (0.296 ± 0.176 MPa, mean ± S.D.). Those values are in good agreement with the values of ultimate tensile stress obtained by other studies. Lawlor et al. showed that the ultimate tensile stress of fresh carotid artery plaques from circumferential direction using uniaxial tensile test was in the range between 0.131 and 0.779 MPa (0.367 ± 0.213 MPa, mean ± S.D.) [10] . Teng et al. performed uniaxial tensile tests on human carotid fibrous cap in the circumferential direction and obtained the ultimate strength of fibrous cap was 0.158 [0.072, 0.259] MPa [52]. Holzapfel et al. reported that the ultimate tensile stress of human carotid fibrous cap in the circumferential direction was 0.255±0.08 MPa [15].

The values of critical energy release rate across the thickness of fibrous cap obtained in the current study ranged from 0.085 to 1.428 N/mm, with a mean value 0.533 ± 0.449 N/mm, which show a large variation from specimen to specimen or from position to position of each specimen. Under the same loading condition, the fibrous cap with lower energy release rate value proned to break, or the positions of one fibrous cap with lower Gc started to tear first. From previous studies of human fibrous cap delamination from the underlying plaque tissue, the critical energy release rate ranged from 0.132 to 0.695 N/mm, with a mean of 0.254±0.155 N/mm [16]. The values across the thickness of fibrous cap are larger than that between the fibrous cap and underlying tissue in the limit studies. Thus, the critical energy release rate has a large effect on failure mechanism of the atherosclerotic plaque. On one hand, the damage and dissection propagation will occur at the interface between fibrous cap and underlying tissue. On the other hand, the fibrous cap will prone to tearing and breakage across the thickness and the whole plaque tissue will exposed to the blood vessel when the fibrous cap is weaker.

There are several study limitations that should be considered in the interpretation of the obtained results. Firstly, the geometry modeling was built from images of front view and side view, of which the thicknesses were assumed constant along the direction along width of the sample. Secondly, the samples under uniaxial tensile test did not undergo pure tensile deformation since the shearing forces may occur inside the samples with large width-length ratio. Thirdly, the material parameter values identified from this study were from a single stretch ratio which may not well characterize the mechanical behavior of the tearing test of arterial tissue. Since the stretch ratio of the tearing test are larger than that from the uniaxial tensile test. Fourthly, the material parameter values were obtained from a local optimal process, so other combinations of different parameter values can characterize a similar mechanical response as the set of parameters in this study.

# 7. CONCLUSIONS

In this study, an inverse analysis method and a finite element based modeling approach for simulating human fibrous cap tearing event were developed and demonstrated in order to identify the cohesive interface parameter values. Simulations of human fibrous cap tearing experiments were carried out, in which the HGO model for the bulk material behavior and the CZM for the tearing behavior along the tearing path were adopted. By implementation of the inverse modeling of the uniaxial tensile tests of human fibrous cap, the HGO parameter values were obtained and used as the input values for the bulk material model to model the fibrous cap tearing process. With the same inverse analysis method, the cohesive interfacial strength and critical energy release rate across the thickness of fibrous cap were quantifed.

Comparisons of simulation predictions of the load-displacement curve from inverse modeling with experimental measurements reveal that the simulation predictions were able to characterize the critical features of the load-displacement curve from the human fibrous cap tearing tests. The results of this study provide a finite element based inverse analysis method to obtain the material parameter values from the simulations of arterial tissue failure events using CZM approach. Furthermore, this study provides a perspective to use finite element method to predict the atherosclerotic plaque rupture and a solid basis for medical advances in intervention and prevention of life-threatening event.

**ACKNOWLEDGEMENTS**

The authors gratefully acknowledge the sponsorship of NSF (award # CMMI-1200358) and this work was partially supported by a SPARC Graduate Research Grant from the Office of the Vice President for Research at the University of South Carolina.